\pdfoutput=1
\documentclass[sigconf]{acmart}

\usepackage{tikz}
\usetikzlibrary{calc}
\usetikzlibrary{shapes.geometric, arrows, calc}
\tikzstyle{node} = [circle, draw, inner sep=2pt, fill=black]
\tikzstyle{node_white} = [circle, draw, inner sep=2pt, fill=white]
\tikzstyle{node_gray} = [circle, draw, inner sep=2pt, fill=gray]
\tikzstyle{node_red} = [circle, draw, inner sep=2pt, color=red, fill=red]
\tikzstyle{node_red_white} = [circle, draw, inner sep=2pt, color=red, fill=white]
\tikzstyle{node_blue} = [circle, draw, inner sep=2pt, color=blue, fill=blue]
\tikzstyle{block} = [draw, text width=2cm, minimum height=1cm, align=center, fill=black, text=white]
\tikzstyle{text_block} = [align=center]

\usepackage{pgfplots}
\pgfplotsset{width=8cm, compat=1.9}

\AtBeginDocument{%
  }

\setcopyright{acmlicensed}
\copyrightyear{2024}
\acmYear{2024}
\acmDOI{10.48550/arXiv.2410.22382}

\acmConference[ICAIF '24]{5th ACM International Conference on AI in Finance}{November 14--16,
  2024}{Brooklyn, NY}
\acmBooktitle{Workshop on Explainable AI in Finance (XAIFIN2024),
  November 15, 2024, Brooklyn, NY}
\acmISBN{978-1-4503-XXXX-X/18/06}

\begin{document}

\title[Debiasing Alternative Data for Credit Underwriting Using Causal Inference]{Debiasing Alternative Data for Credit Underwriting\\Using Causal Inference}

\author{Chris Lam}
\affiliation{%
  \institution{Epistamai}
  \city{Apex}
  \state{NC}
  \country{USA}}
\email{chris@epistam.ai}

\renewcommand{\shortauthors}{Lam}

\begin{abstract}
  Alternative data provides valuable insights for lenders to evaluate a borrower's creditworthiness, which could help expand credit access to underserved groups and lower costs for borrowers. But some forms of alternative data have historically been excluded from credit underwriting because it could act as an illegal proxy for a protected class like race or gender, causing redlining. We propose a method for applying causal inference to a supervised machine learning model to debias alternative data so that it might be used for credit underwriting. We demonstrate how our algorithm can be used against a public credit dataset to improve model accuracy across different racial groups, while providing theoretically robust nondiscrimination guarantees.
\end{abstract}



\received{2 October 2024}

\maketitle

\section{Introduction}
The financial services industry has for years discussed the potential of using alternative data sources to improve lenders' ability to accurately assess the creditworthiness of borrowers \cite{Kreiswirth, FICO2023, yingleitoh, terribradford}. Traditional credit scores like FICO have historically relied on factors like payment history, amounts owed, length of credit history, and recent credit inquiries to generate their models \cite{FICO2024}. More recently, alternative data sources like transactions or cash flow (e.g. bank accounts), bill payments (e.g. rent and utilities), and income or assets (e.g. employment history and property ownership) are also being utilized in some credit scoring models. By integrating more data into these models, this can lead to improved accuracy that could help lenders expand credit access to underserved borrowers such as credit invisible and thin file loan applicants \cite{cfpb_credit_invisible}, while enabling more borrowers to be approved or to qualify for lower interest rates \cite{NBERw29840, FinRegLab2020, Turner}.

AI researchers have identified additional sources of alternative data that could aid credit underwriting but may not currently be in use by the industry. For example, Berg identified digital footprints, such as whether borrowers browsed online in incognito mode or used an anonymous email address, as being more likely to default \cite{Berg2020-am}. Lee identified certain grocery shopping behaviors, like whether someone purchases cigarettes or processed meat, as also being less likely to pay back a loan \cite{lee2021}. Finally, Netzer identified certain words or phrases that were used on a credit application, like ``please,'' ``promise,'' and ``thank you,'' as potential red flags \cite{Netzer2019-uh}.

But these non-financial sources of alternative data may entail greater regulatory risk. Such data may act as an illegal proxy for a protected class like race or gender, leading to potentially unlawful discrimination \cite{Barocas2016-pn, prince, tschantz2022proxy}. Many FinTechs would like to leverage such alternative data with machine learning algorithms to improve model accuracy, but don't know how to do so without introducing unwanted bias or illegal discrimination. A large part of this problem has to do with the fairness through unawareness approach that is adopted by regulations such as the Equal Credit Opportunity Act Regulation B in the US, which several AI researchers have raised concerns about \cite{Barocas2016-pn, dwork2011fairnessawareness, hardtbigdata, 10.1145/1401890.1401959, kleinberg2018}.

A small community of AI researchers have proposed using causal models as a potential solution to the fairness and machine learning problem \cite{Chiappa2019, nilforoshan2022causal, plecko2022causal, barocas-hardt-narayanan, Chockler_Halpern_2022, kilbertus2018avoidingdiscriminationcausalreasoning, Weerts_2024}. In particular, techniques proposed by Pearl using causal Bayesian networks (CBNs) may offer a particularly promising solution \cite{Pearl2016-jx, Pearl2020-yt}. But to use these methods require building proper causal models of the fairness problem, which is not easy and requires significant domain expertise.

This paper will discuss how to build causal fairness models for credit underwriting, showing a fundamental relationship between supervised machine learning and causal inference. We will then discuss how to use those models to identify an algorithm to debias alternative data so that it would not act as an illegal proxy. Finally, we will show the results of this algorithm against a public dataset (i.e. the National Survey of Mortgage Originations).

\section{Building a causal model of fairness}

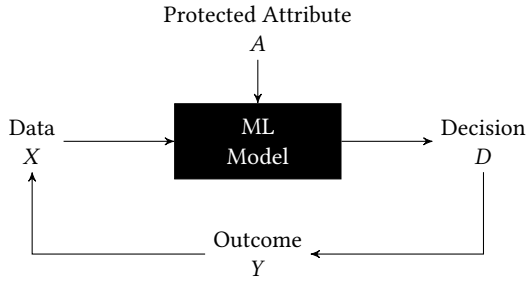
\begin{figure}[ht]
    \centering
    \begin{tikzpicture}[->, >=stealth']
        \node[block] (box) {ML \\Model};
        \node[text_block, left of=box, xshift=-2cm] (x) {Data\\$X$};
        \node[text_block, right of=box, xshift=2cm] (d) {Decision\\$D$};
        \node[text_block, above of=box, yshift=0.5cm] (a) {Protected Attribute\\$A$};
        \node[text_block, below of=box, yshift=-0.5cm] (y) {Outcome\\$Y$};
        \draw[->] (x) -- (box);
        \draw[->] (box) -- (d);
        \draw[->] (a) -- (box);
        \draw (d) |- (y);
        \draw (y) -| (x);
    \end{tikzpicture}
    \caption{Block diagram representation of the supervised machine learning problem}
    \Description{A black box ML model with data $X$ as input and decision $D$ as output. A protected attribute $A$ is also fed into the ML model. The decision $D$ leads to an outcome $Y$, which is then fed back into data $X$.}
    \label{fig:box}
\end{figure}

We begin with a block diagram representation of the supervised machine learning problem as shown in Figure \ref{fig:box}. In the center is a black box which represents a machine learning model such as a neural network. The black box inputs data $X$ and outputs decision $D$. The decision $D$ leads to a future outcome $Y$, which in turn becomes future data $X$. The key question is whether a protected attribute $A$ should be used as an input to the black box model, and if so, how should it be used?

In credit underwriting, data $X$ could represent traditional or alternative data, such as credit bureau data that captures past outcomes $Y$. The decision $D$ could represent whether a loan is approved and its interest rate. The outcome $Y$ could represent whether a borrower defaulted. The protected attribute $A$ could represent race or gender, for example.

We now want to transform this block diagram representation into a causal Bayesian network (CBN). We perform this transformation through a series of steps. Note that due to the directed acyclic graph constraint of CBNs, we do not complete the feedback loop between outcome $Y$ and data $X$.

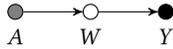
\begin{figure}[ht]
    \centering
    \begin{tikzpicture}[->, >=stealth', node distance=1cm]
	\node[node_gray, label=below:$A$] (A) {};
	\node[node_white, label=below:$W$] (W) [right of=A] {};
	\node[node, label=below:$Y$] (Y) [right of=W] {};
	\path
		(A) edge node {} (W)
		(W) edge node {} (Y);
    \end{tikzpicture}
    \caption{A causal Bayesian network (CBN) showing the relationship between a protected attribute $A$, a mediator $W$, and an outcome $Y$}
    \Description{A causal Bayesian network with protected attribute $A$ causing latent mediator $W$, which in turn causes outcome $Y$.}
    \label{fig:awy}
\end{figure}

In Figure \ref{fig:awy}, we have a three node CBN where the protected attribute $A$ causes the mediator $W$, which in turn causes the outcome $Y$. The protected attribute $A$ is colored gray because it may or may not be observable. If not, it could be imputed using a technique like Bayesian Improved Surname Geocoding (BISG) \cite{cfpb_bisg}. The mediator $W$ represents creditworthiness and is colored white because it is a latent variable (i.e. it is not fully observable) \cite{evans}. On the other hand, the outcome $Y$ is colored black because it is observable. The outcome $Y$ is also independent of the protected attribute $A$ given the mediator $W$: $Y \perp\!\!\!\perp A|W$. In other words, creditworthiness completely explains away the relationship between the protected attribute and whether a borrower would default.

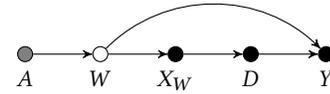
\begin{figure}[ht]
    \centering
    \begin{tikzpicture}[->, >=stealth', node distance=1cm]
	\node[node_gray, label=below:$A$] (A) {};
	\node[node_white, label=below:$W$] (W) [right of=A] {};
	\node[node, label=below:$X_W$] (X_W) [right of=W] {};
	\node[node, label=below:$D$] (D) [right of=X_W] {};
	\node[node, label=below:$Y$] (Y) [right of=D] {};
	\path
		(A) edge node {} (W)
		(W) edge node {} (X_W)
		(X_W) edge node {} (D)
		(D) edge node {} (Y)
		(W) edge[bend left=45] node {} (Y);
    \end{tikzpicture}
    \caption{A CBN with traditional data $X_W$ and a decision $D$}
    \Description{A causal Bayesian network where data $X_W$ is a measure of mediator $W$, which causes decision $D$. This in turn causes outcome $Y$.}
    \label{fig:awxyy}
\end{figure}

In Figure \ref{fig:awxyy}, we add in a few more nodes. Traditional data $X_W$ is a partial and imperfect measure of the latent mediator $W$ that is used to make a decision $D$. For example, we can set $D$ as
\begin{equation}\label{DYXW}
    D=P(Y=y|X_W=x_W)<t
\end{equation}
where $t$ is some threshold for predicted loan default. If a borrower is predicted to be below the threshold $t$, then he or she may be approved for a loan. Alternatively, a scorecard could be used with various thresholds to determine an interest rate. The machine learning model is embedded inside the decision node $D$, which also has a causal effect on the outcome $Y$. For example, borrowers cannot default if they were never approved for a loan.

The graph in Figure \ref{fig:awxyy} represents how traditional credit scoring is performed today. Traditional data $X_W$ does not act as a proxy for the protected attribute $A$ because it is a measure of the mediator $W$ (i.e. creditworthiness), which acts as a confounding variable through the fork $X_W \leftarrow W \rightarrow Y$. In these causal models, we would also consider alternative data in use by the industry today (e.g. cash flow or utilities data) as traditional data $X_W$ because they do not proxy for the protected attribute $A$. In this graph, the protected attribute $A$ has no influence on the decision $D$. This is also known among the AI research community as fairness through unawareness.

One may view Figure \ref{fig:awy} as a \textit{world model}. Figure \ref{fig:awxyy} shows how to overlay a \textit{machine learning model} on top of the world model through the addition of the variables $X_W$ and $D$. One may also consider the machine learning model to be a surrogate of the world model. That is, the data $X_W$ acts as a surrogate for the mediator $W$ and the decision $D$ acts as a surrogate for the outcome $Y$. For those who are familiar with Pearl's three-layer causal hierarchy \cite{Pearl2020-yt}, the machine learning model is operating at the first layer (association), while the world model is operating at the second layer (intervention) and third layer (counterfactuals).

\section{Defining unlawful discrimination using causal models}

There are two legal doctrines for unlawful discrimination in credit underwriting \cite{fredman}. In the US, these are called disparate treatment and disparate impact. In the EU, these are called direct discrimination and indirect discrimination. This paper will focus on US fair lending discrimination. 

According to the Interagency Fair Lending Examination Procedures manual \cite{ffiec}, disparate treatment occurs when a lender explicitly considers prohibited factors (overt evidence) or by differences in treatment that are not fully explained by legitimate nondiscriminatory factors (comparative evidence). On the other hand, disparate impact occurs when a formally neutral policy disproportionately excludes or burdens certain persons on a prohibited basis.

\begin{figure}[ht]
    \centering
    \begin{tikzpicture}[->, >=stealth', node distance=1cm]
	\node[node_gray, label=below:$A$] (A) {};
	\node[node_white, label=below:$W$] (W) [right of=A] {};
	\node[node, label=below:$X_W$] (X_W) [right of=W] {};
	\node[node_red, label=below:$D$] (D) [right of=X_W] {};
	\node[node_red, label=below:$Y$] (Y) [right of=D] {};
	\path
		(A) edge node {} (W)
		(W) edge node {} (X_W)
		(X_W) edge node {} (D)
		(D) edge[color=red] node {} (Y)
		(W) edge[bend left=45] node {} (Y)
            (A) edge[bend right=45, color=red] node {} (D);
    \end{tikzpicture}
    \caption{Overt discrimination}
    \Description{A red edge from the protected attribute $A$ to decision $D$ indicates bias due to overt discrimination. This also leads to a red edge from decision $D$ to outcome $Y$, also indicating bias.}
    \label{fig:overt}
\end{figure}
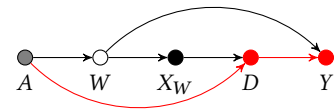

We can model disparate treatment as a form of overt discrimination using the graph in Figure \ref{fig:overt}. The protected attribute $A$ is explicitly used to influence a decision $D$, which causes both the decision $D$ and the outcome $Y$ to become biased. We can visualize this bias using red nodes and edges. For example, a lender that uses knowledge of a borrower being Black as a basis for denying a loan would cause overt discrimination that would lead to disparate treatment. In causal terms, we would say that the protected attribute $A$ has a negative direct effect on the decision $D$. This is why the financial services industry generally uses fairness through unawareness, as shown in Figure \ref{fig:awxyy}.

We can model disparate impact using a three step process. The first step for establishing disparate impact is to identify whether a disparity exists on a prohibited basis. If so, the next step is to determine whether the policy is justified by a ``business necessity.'' Finally, a lender could still be in violation if an alternative policy or practice could achieve a less discriminatory effect.

\begin{figure}[ht]
    \centering
    \begin{tikzpicture}[->, >=stealth', node distance=1cm]
	\node[node_gray, label=below:$A$] (A) {};
	\node[node_red_white, label=below:$W$] (W) [right of=A] {};
	\node[node_red, label=below:$X_W$] (X_W) [right of=W] {};
	\node[node_red, label=below:$D$] (D) [right of=X_W] {};
	\node[node_red, label=below:$Y$] (Y) [right of=D] {};
	\path
		(A) edge[color=red] node {} (W)
		(W) edge[color=red] node {} (X_W)
		(X_W) edge[color=red] node {} (D)
		(D) edge[color=red] node {} (Y)
		(W) edge[bend left=45, color=red] node {} (Y);
    \end{tikzpicture}
    \caption{Covert discrimination}
    \Description{A causal Bayesian network where all of the nodes and edges are red to indicate bias due to covert discrimination.}
    \label{fig:covert}
\end{figure}
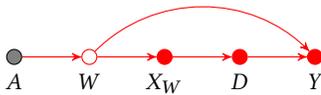

The first step of disparate impact could be modeled as a form of covert discrimination using the graph in Figure \ref{fig:covert}. The protected attribute $A$ is not explicitly used to influence a decision $D$, which makes the policy formally neutral. That is, the protected attribute $A$ has no direct effect on the decision $D$. However, due to biases that may exist from the use of traditional data $X_W$ alone, it could disproportionately burden or exclude certain persons on a prohibited basis. Such biases may be due to historical, structural, and systemic discrimination for example. Toh argues that the use of traditional credit scoring models may disproportionately punish consumers from economically disadvantaged groups including minorities, who are more likely to be credit invisible or have thin files \cite{yingleitoh}. Here we can visualize bias across the entire graph through red nodes and edges. In causal terms, we would say that the protected attribute $A$ has a negative indirect effect on the decision $D$. This is why the financial services industry's approach of fairness through unawareness, as shown in Figure \ref{fig:awxyy}, may be insufficient for preventing disparate impact.

The following section will discuss how the use of alternative data for credit scoring may help to satisfy the second and third steps of disparate impact. That is, alternative data may be able to improve the accuracy of a credit scoring model, which is necessary for satisfying the business necessity requirement. Finally, it may also be able to improve fairness by picking up additional signals for economically disadvantaged groups to prove their creditworthiness, thus leading to a less discriminatory credit scoring model.

\section{Incorporating alternative data into a causal model}

One of the primary goals of a data scientist is to build the most accurate model. For example, if we had perfect information about a borrower's creditworthiness and combined it with a perfect model of the world, we should be able predict with perfect accuracy whether a borrower would default on a loan. A perfectly accurate model would help lenders because they would be able to make loans to all creditworthy borrowers in the marketplace without taking any losses from defaults. This would also help borrowers in a competitive market for credit as they would be charged lower risk premiums and never have to deal with the consequences of defaulting on a loan. However, this goal is not fully achievable in practice. One of the reasons is due to limitations in our ability to collect traditional data $X_W$ to measure the mediator $W$ due to practical, legal, or ethical reasons.

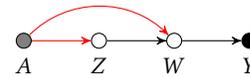
\begin{figure}[ht]
    \centering
    \begin{tikzpicture}[->, >=stealth', node distance=1cm]
	\node[node_gray, label=below:$A$] (A) {};
	\node[node_white, label=below:$Z$] (Z) [right of=A] {};
	\node[node_white, label=below:$W$] (W) [right of=Z] {};
	\node[node, label=below:$Y$] (Y) [right of=W] {};
	\path
		(A) edge[color=red] node {} (Z)
		(Z) edge node {} (W)
		(W) edge node {} (Y)
		(A) edge[bend left=45, color=red] node {} (W);
    \end{tikzpicture}
    \caption{A CBN with demographic variable $Z$}
    \Description{Demographic variable $Z$ partially explains away the causal effect between protected attribute $A$ and mediator $W$. The red edges indicate bias due to the spurious proxy effect that $Z$ has on $W$ through $A$.}
    \label{fig:azwy}
\end{figure}

Since we can never fully capture the mediator $W$ using traditional data $X_W$, there is interest in the industry to use alternative data to gain a better picture of a borrower's creditworthiness. But certain types of alternative data, such as zip code or education, may act a proxy for a protected class. These types of alternative data are modeled in Figure \ref{fig:azwy} as demographic variables $Z$, which is a modified version of Figure \ref{fig:awy}. Unlike the mediator $W$, demographic variables $Z$ do not completely explain away the causal effect between the protected attribute $A$ and the outcome $Y$. As a result, the protected attribute $A$ may act as a confounder variable for demographic variables $Z$. Note that both the mediator $W$ and demographic variables $Z$ are latent and thus colored white.

Demographic variables $Z$ are correlated with the creditworthiness mediator $W$ through two paths. The first is a direct causal path $Z \rightarrow W$, which should be considered a legal path. The second is a spurious non-causal path $Z \leftarrow A \rightarrow W$, which should be considered an illegal path through the protected attribute $A$. This spurious path is colored red in Figure \ref{fig:azwy}.

\begin{figure}[ht]
    \centering
    \begin{tikzpicture}[->, >=stealth', node distance=1cm]
	\node[node_gray, label=below:$A$] (A) {};
        \node[node_white, label=below:$P$] (P) [below right of=A] {};
	\node[node_white, label=below:$W$] (W) [right of=A] {};
	\node[node, label=below:$Y$] (Y) [right of=W] {};
	\path
		(A) edge[color=red] node {} (W)
		(A) edge[color=red] node {} (P)
		(W) edge node {} (Y);
    \end{tikzpicture}
    \caption{A CBN with proxy variable $P$}
    \Description{Proxy variable $P$ has no causal effect on mediator $W$. The red edges indicate bias due to the spurious proxy effect that $P$ has on $W$ through $A$.}
    \label{fig:apzwy}
\end{figure}
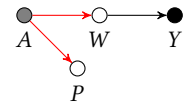

We also need to be careful not to include any variables that have no causal relationship to the creditworthiness mediator $W$. Figure \ref{fig:apzwy} shows a proxy variable $P$, which is only correlated with the mediator $W$ through the spurious non-causal path $P \leftarrow A \rightarrow W$. This should also be considered an illegal path through the protected attribute $A$.

\begin{figure}[ht]
    \centering
    \begin{tikzpicture}[->, >=stealth', node distance=1cm]
	\node[node_gray, label=below:$A$] (A) {};
	\node[node_white, label=below:$Z$] (Z) [right of=A] {};
	\node[node, label=below:$X_Z$] (X_Z) [right of=Z] {};
	\node[node_white, label=below:$W$] (W) [right of=X_Z] {};
	\node[node, label=below:$X_W$] (X_W) [right of=W] {};
	\node[node, label=below:$D$, color=red] (D) [right of=X_W] {};
	\node[node, label=below:$Y$, color=red] (Y) [right of=D] {};
	\path
		(A) edge node {} (Z)
		(Z) edge node {} (X_Z)
		(Z) edge[bend left=45] node {} (W)
		(W) edge node {} (X_W)
		(X_W) edge node {} (D)
		(D) edge[color=red] node {} (Y)
		(A) edge[bend left=45] node {} (W)
		(W) edge[bend left=45] node {} (Y)
            (X_Z) edge[bend right=45, color=red] node {} (D);
    \end{tikzpicture}
    \caption{A CBN with alternative data $X_Z$}
    \Description{A red edge from alternative data $X_Z$ to decision $D$ indicates bias due to redlining and proxy discrimination.}
    \label{fig:azxwxyy}
\end{figure}

By modifying Figure \ref{fig:awxyy} to include alternative data $X_Z$, we get Figure \ref{fig:azxwxyy}. Notice that the direct use of alternative data $X_Z$ might trigger proxy discrimination and lead to overt discrimination in the decision $D$, similar to the overt discrimination caused by the direct use of the protected attribute $A$ in Figure \ref{fig:overt}. This is because as shown in Figure \ref{fig:azwy}, demographic variable $Z$ is spuriously correlated with the mediator $W$ through the protected attribute $A$. That is, the machine learning classifier has no way to isolate the direct causal effect $Z \rightarrow W$ from the spurious non-causal effect $Z \leftarrow A \rightarrow W$.

\begin{figure}[ht]
    \centering
    \begin{tikzpicture}[->, >=stealth', node distance=1cm]
	\node[node_gray, label=below:$A$] (A) {};
	\node[node_white, label=below:$Z$] (Z) [right of=A] {};
	\node[node, label=below:$X_Z$] (X_Z) [right of=Z] {};
	\node[node_white, label=below:$W$] (W) [right of=X_Z] {};
	\node[node, label=below:$X_W$] (X_W) [right of=W] {};
	\node[node, label=below:$D$] (D) [right of=X_W] {};
	\node[node, label=left:${A=a'}$] (Ap) [above of=X_W] {};
	\node[node, label=below:$Y$] (Y) [right of=D] {};
	\path
		(A) edge node {} (Z)
		(Z) edge node {} (X_Z)
		(Z) edge[bend left=45] node {} (W)
		(W) edge node {} (X_W)
		(X_W) edge node {} (D)
		(D) edge node {} (Y)
		(A) edge[bend left=45] node {} (W)
		(W) edge[bend left=45] node {} (Y)
            (X_Z) edge[bend right=45, color=red] node {} (D)
            (Ap) edge[color=blue] node {} (D);
    \end{tikzpicture}
    \caption{A CBN with alternative data $X_Z$}
    \Description{A blue edge from $A=a'$ indicates a bias correction from the use of $X_Z$ on the decision $D$.}
    \label{fig:azxwxyy2}
\end{figure}

But we can correct for this problem by training a classifier for decision $D$ with the protected attribute $A$. Then during inference, we make an intervention where we set a default value for the protected attribute to $A=a'$ in order to prevent alternative data $X_Z$ from acting as a proxy for a protected attribute $A$, as shown in Figure \ref{fig:azxwxyy2}. For example, a Black loan applicant using only traditional data $X_W$ would be scored using fairness through unawareness. However, that same Black loan applicant being scored using alternative data $X_Z$ would need some control for the proxy effect, which we could accomplish by treating every loan applicant regardless of race as White. This should satisfy the requirements for disparate treatment, since everyone is treated the same regardless of their race.

More specifically, we modify Equation \ref{DYXW} and set D as
\begin{equation}
    D=P(Y=y|do(A=a', X_Z=x_Z, X_W=x_W))<t
\end{equation}

To replace the do-operator, we notice that since we already conditioned on $A$, all of the relevant backdoor paths that could create a proxy for a protected class are blocked. Thus, all we have to do is simply control for those variables
\begin{equation}
    D=P(Y=y|A=a', X_Z=x_Z, X_W=x_W)<t
\end{equation}

Note that this would not be considered a full backdoor adjustment as there could be additional confounding variables for the demographic variable $Z$ besides the protected attributes $A$. That is, the protected attributes $A$ are a subset of Pearl's backdoor criterion. But for the purposes of preventing proxy discrimination and disparate treatment, controlling for only the protected attributes $A$ should be sufficient.

In addition, this approach also corrects for disparate impact by allowing us to use alternative data to expand credit to economically disadvantaged groups. Therefore, unlike Figure \ref{fig:covert} which shows bias in the data from the exclusive use of traditional data $X_W$, Figure \ref{fig:azxwxyy2} does not show bias because we also include alternative data $X_Z$. This additional data will improve the accuracy of the classifier in decision $D$, which satisfies step 2 of disparate impact. Finally, it would help us to identify additional borrowers from economically disadvantaged groups who are more creditworthy than a traditional credit score would suggest.

\section{Leveraging causal inference to debias alternative data}

Now that we have causal graphs for modeling fairness and discrimination in the credit underwriting problem, we will now dive deeper into explaining how we can use Pearl's tools of causal inference to make interventions into a machine learning model \cite{10.1145/3241036}. This provides the foundation for a causal approach to data science that could be called ``model-based'' supervised learning, which is in contrast to the ``model-free'' supervised learning that the industry commonly uses today. By using a CBN as a higher level abstraction of the supervised ML problem, it acts as a type of world model that provides a high-level specification for performing data science.

The data science process requires performing a series of steps, including feature selection, data preparation, training, inference, and evaluation. While each of these steps are already intuitively performed by a data scientist today, they can be more systematically augmented using causal inference.

The first step is feature selection, which is a type of \textit{mediation analysis}. A data scientist must select features that have a causal relationship to a borrower's creditworthiness (or as Evans points out, a ``nexus to creditworthiness'' \cite{evans}). This can be done using prior knowledge of known causal effects in the data. In the financial services industry, some professionals refer to creditworthiness as the 5 C's of credit: character, capacity, capital, collateral and conditions \cite{Peterdy, Segal}. Traditional data and some forms of financial alternative data (e.g. bank cash flow, rent and utility payments) may be considered measures of the creditworthiness mediator $W$, and are therefore labeled as traditional data $X_W$. Non-financial forms of alternative data $X_Z$ (e.g. zip code and education) may be considered measures of demographic variables $Z$, which have both a direct causal effect on the creditworthiness mediator $W$ and a spurious non-causal effect through the confounding protected attribute $A$, creating an illegal proxy effect. This is in contrast to data that have no causal effect on the creditworthiness mediator $W$, which must be excluded from the model. They instead may only be spuriously correlated with outcome $Y$ through the confounding protected attribute $A$ (e.g. name, hair length, or music preference).

This is followed by the data preparation step. A data scientist needs to clean up the data such as by removing or replacing values. During this step, a data scientist should perform \textit{overlap testing}. That is, we need to ensure that all of the values for each feature have an overlap with the values of a protected class. If there is no overlap, then we would have a \textit{positivity violation}. This is especially important for alternative data $X_Z$. An example of a positivity violation may be using a historically Black college or university (i.e. HBCU) for education. Very few if any Whites attend an HBCU, making it impossible for a classifier to isolate the direct causal effect of attending an HBCU on the creditworthiness mediator $W$ from the spurious non-causal effect of the confounding protected attribute for race $A$. When such a positivity violation occurs, a data scientist would need to remove or replace the value for the feature. Note that you do not necessarily have to drop the entire feature, just individual values where the positivity violation exists.

We can think of the training step as a type of \textit{backdoor adjustment}. We train a classifier using the protected attribute $A$ in order to isolate the direct causal effect of demographic variable $Z$ on the creditworthiness mediator $W$ (i.e. $Z \rightarrow W$) from the spurious non-causal effect through the confounding protected attribute $A$ (i.e. $Z \leftarrow A \rightarrow W$). This has the effect of debiasing alternative data $X_Z$ by closing off the backdoor path that could cause it to act as an illegal proxy for the protected attribute $A$.

During the inference step, we perform our discrimination testing by making an \textit{intervention} into the model. For a given loan applicant, we would perform a paired test using both the default values and the actual values for the protected attribute $A$. As an example, we may use White as a default value for race and male as a default value for gender. More specifically, we would be performing Pearl's \textit{do-intervention} to replace the protected attribute $A$'s direct effect on the decision $D$, such as $P(Y=y|do(A=a'), do(X_Z=x_Z), do(X_W=x_W))$. This has the causal effect of closing the backdoor effect of the feature $X_Z$ through the confounding protected attribute $A$. For example, a zip code where the population is majority Black cannot act as a proxy for a borrower being Black if the classifier was told to treat that borrower as a White person living in that zip code. This assumes that we have enough examples of White people living in that zip code to train on so that the classifier can isolate its causal effect from its spurious effect. This will allow the classifier to use zip code to capture local economic conditions that are independent of race.

Finally, during the evaluation step we perform a \textit{counterfactual analysis}. We compare the decision $D$ that is derived from a traditional credit score (as shown in Figure \ref{fig:awxyy}) to an alternative credit score (as shown in Figure \ref{fig:azxwxyy2}). This will allow us to verify that the alternative credit scoring model is less discriminatory across all protected classes. We may consider this as another approach to using counterfactuals for fairness \cite{kusner2018counterfactual, kilbertus2019sensitivity, chiappa2018pathspecificcounterfactualfairness, nabi2018fairinferenceoutcomes}.

\section{Testing the Algorithm}

Our debiasing algorithm was tested against a public dataset called the National Survey of Mortgage Originations (NSMO). This survey is administered by the Federal Housing Finance Agency (FHFA) and the CFPB. It is conducted quarterly to collect voluntary feedback directly from mortgage borrowers about their experiences with obtaining a mortgage. While this public dataset is not actual credit bureau data, we used it to simulate that type of data. One of the unique aspects of this public dataset is that it contains both a protected attribute (race) and its outcome (default).

The specific version of the NSMO dataset that we used was \texttt{nsmo\_v41\_1320\_puf.csv}, which was released in March 2023. We generated the target variable using the \texttt{Perf\_Status} questions, marking a loan as having defaulted if it had codes 2 through 9, or at least 60 days past due. We combined the variables X77R and X78R to generate the following classes for race: Non-Hispanic White, Hispanic, Black, Asian, and other. The focus of our analysis for this paper was between Non-Hispanic Whites and Blacks.

The remaining X and Z variables were used to train our models, as well as several supplemental features. Since the focus of our test was on race, we decided to leave in age and sex. However, these could be dropped from our models without loss of generality. We also removed supplemental variables to prevent data leakage. Our source code has been publicly uploaded to Github for researchers who want to replicate our results.\cite{github}

\subsection{Model training}

We trained three models, which we called the awareness, unawareness, and counterfactual models. The awareness model is a ``kitchen-sink model'' that used every feature including race, which yielded the highest accuracy but also caused disparate treatment. The unawareness model dropped race and any features highly correlated with Black applicants. For the purpose of this experiment, we assumed that such features could be considered a form of alternative data that could act as an illegal proxy. This model simulates the fairness through unawareness approach commonly used by the industry today. Finally, the counterfactual model uses all features including race during training but treats all applicants as White during inference, which is equivalent to Figure \ref{fig:azxwxyy2} to prevent disparate treatment. Since it uses alternative data to expand credit access to underserved groups, it would also prevent disparate impact.

The models were built using the Microsoft LightGBM library. We used an 80/20 train/test split with 5 K-fold validation. The evaluation metric was ROC/AUC.

For the unawareness model, we dropped any features that were highly correlated with race, and more specifically with being Black. Any features that had a correlation coefficient less than -0.05 or greater than 0.05 were dropped, for a total of 89 out of 1012 features.

For the two backdoor models, we also did overlap testing to ensure that feature values for all the potential proxy features had overlap with the racial groups. We didn't identify any positivity violations with this dataset. But had we identified such violations, we would have simply dropped the values for those proxy features.

\subsection{Results}

\begin{figure}[ht]
    \centering
    \begin{tikzpicture}
        \begin{axis}[
            enlargelimits=0.15,
            legend style={at={(0.5, -0.15)}, anchor=north, legend columns=-1},
            ylabel={ROC/AUC score},
            symbolic x coords={Awareness, Unawareness, Counterfactual},
            xtick=data
        ]
        \addplot[
            black,
            mark=*,
            nodes near coords,
            every node near coord/.append style={/pgf/number format/.cd, fixed, fixed zerofill, precision=4}
        ] coordinates {(Awareness, 0.8761) (Unawareness, 0.8705) (Counterfactual, 0.8758)};
        \addplot[
            red,
            mark=square*,
            nodes near coords,
            nodes near coords style={anchor=north},
            every node near coord/.append style={/pgf/number format/.cd, fixed, fixed zerofill, precision=4}
        ] coordinates {(Awareness, 0.8728) (Unawareness, 0.8675) (Counterfactual, 0.8728)};
        \addplot[
            blue,
            mark=triangle*,
            nodes near coords,
            every node near coord/.append style={/pgf/number format/.cd, fixed, fixed zerofill, precision=4}
        ] coordinates {(Awareness, 0.8190) (Unawareness, 0.8111) (Counterfactual, 0.8186)};
        \legend{Overall, White, Black}
        \end{axis}
    \end{tikzpicture}
    \caption{ROC/AUC curves}
    \Description{The counterfactual model is able to recover most of the loss in ROC/AUC from the unawareness model.}
    \label{results}
\end{figure}
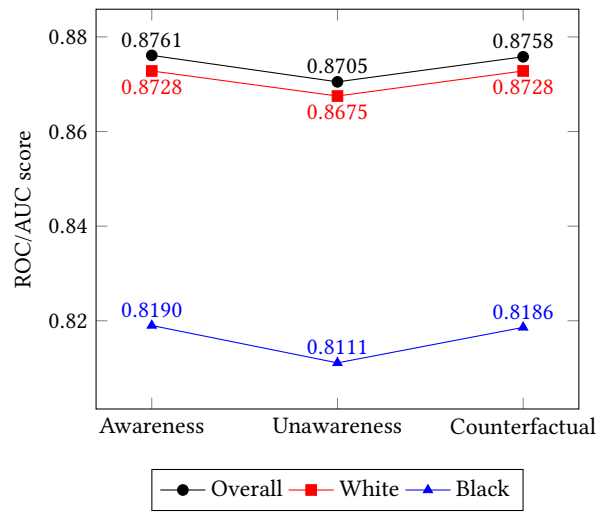

For each model, we ran the tests 30 times using different random seeds in order to prove statistical significance. We then evaluated model performance using the ROC/AUC score. Figure \ref{results} shows the results of each of the four models. 

While there was a significant drop-off in accuracy going from the awareness to unawareness models, most of that dropoff was recovered by moving to the counterfactual model. Note that there was still a very small dropoff in accuracy between the awareness and the counterfactual model, which is due to the cost of mislabeling Black applicants as White. However, that cost was more than offset by the increased use of data to explain away the effect of race.

\section{Conclusion}

Causal inference may provide a more robust set of algorithmic techniques for debiasing alternative data to prevent it from acting as an illegal proxy for a protected attribute. Compared to the statistical or correlational techniques commonly used by the industry today, such as hyperparameter tuning or adversarial debiasing \cite{relman_upstart}, causal techniques may provide a stronger mathematical and theoretical basis for proving that a machine learning algorithm will not cause unlawful discrimination.

Unfortunately, current fair lending regulations like the Equal Credit Opportunity Act (ECOA) in the US may not allow the use of our causal debiasing algorithm at this time. According to ECOA Regulation B, ``a creditor shall not consider race, color, religion, national origin, or sex (or an applicant's or other person's decision not to provide the information) in any aspect of a credit transaction'' (1002.6). Instead, the protected class may only be used for monitoring purposes (1002.13), self-testing and correction (1002.15), or for determining eligibility in a special purpose credit program (1002.8). That is, the protected attribute may only be used to detect discrimination after a decision has been made, as opposed to being used to prevent discrimination before a decision has been made.

We believe that policymakers should reconsider this policy as these techniques provide at least five major advantages. First, they directly address the fairness through unawareness concerns that the AI researchers have made over the last several years. These techniques provide robust guardrails that allow us to make full use of data that has a causal relationship or nexus to creditworthiness. Second, they could be used to solve the proxy discrimination problem in alternative data, expanding the amount of data that lenders could use to assess a borrower's creditworthiness. Third, this broader use of alternative data could help expand credit access to underserved groups, including credit invisible and thin file borrowers. Fourth, they enable the development of less discriminatory models for satisfying the third step of disparate impact. Finally, they can help eliminate the need for regulatory sandboxes by providing more transparent tools to audit and correct for bias and discrimination in machine learning models. These causal techniques may provide a means for modernizing our regulations, enabling a more equitable and inclusive financial system that would benefit both lenders and borrowers.

\bibliographystyle{ACM-Reference-Format}
\bibliography{icaif}

\end{document}